

\magnification\magstephalf

\font\rfont=cmr7 at 8.4 true pt
\def\ref#1{$^{\hbox{\rfont {[#1]}}}$}


\font\fourteenbf=cmbx12 scaled\magstep1

\font\twelverm=cmr12


\def\pmb#1{\setbox0=\hbox{#1}
 \kern.05em\copy0\kern-\wd0 \kern-.025em\raise.0433em\box0 }

\def \half {{\scriptstyle {1 \over 2}}}

 %

\def \i {\item}

\parskip=6pt
\parindent=0pt
\hsize=17truecm\hoffset=-5truemm
\voffset=-1truecm\vsize=25.5truecm
\def\footnoterule{\kern-3pt
\hrule width 17truecm \kern 2.6pt}

{\nopagenumbers
{\twelverm
\rightline{CERN-TH-6431/92}

\rightline{M/C-TH 92-09}
\vskip 2truein
\centerline{{\fourteenbf UNUSUAL HIGH  p$_T$ JET EVENTS AT HERA}}
\bigskip
\bigskip
\centerline{A Donnachie}
\centerline{Department of Theoretical Physics}
\centerline{University of Manchester}
\bigskip
\centerline{P V Landshof{}f}
\centerline{CERN, Geneva\footnote{$^{\dag}$}{On leave of absence from
DAMTP, University of Cambridge}}
\vskip 1truein
{\bf Abstract}
\bigskip
We calculate the cross-section for events at HERA where the proton
loses only a minute fraction of its initial energy, all of which
goes into producing a single pair of transverse jets.
\vskip 7truecm
{CERN-TH-6431/92}\hfill\break
{M/C-TH 92-09}

March 1992
\vfill\eject
\vfill\eject
}}
\pageno=1
\bigskip
\bigskip
High-energy long-range strong interactions are generated by pomeron
exchange. The pomeron has been shown experimentally to be almost a
real object: for example, it can hit a hadron hard and knock nearly
all the energy of the resulting fragments almost entirely into the
forward direction\ref{1,2}. There is now a large body of data which show that
in many ways the pomeron is similar to a hadron or a photon in its
effects\ref{2,3}. In order to understand the pomeron better, it is
important to measure its structure function\ref{4}.

The UA8 experiment\ref{1,5} at the CERN collider has done this by studying
 high-$p_T$
jet production
in diffractive $\bar pp$ collisions. The conclusion is that the  shape
of the structure function in the Bjorken variable $x$ fits quite well to
the form
$$
Cx(1-x)
\eqno(1)
$$
but the experiment is not able to determine whether the structure is
predominantly of quarks or of gluons. We have argued\ref{6}
that in fact the structure
function is likely to be predominantly composed of quarks, and have obtained
the shape (1) with the coefficient $C$ a little less than 0.2 for each light
quark or antiquark. The UA8 results\ref{5} are consistent with such a
normalisation, which is an order of magnitude smaller than some other
predictions\ref{7}.

HERA will be able to measure the pomeron's quark  structure function directly,
by determining in
what fraction of virtual-photon inclusive events the initial proton
loses only a small fraction of its energy\ref{4,6}.
But it will be useful also to repeat at HERA the UA8 experiment:
namely, to study high-$p_T$ jet
production in events where the initial proton scarcely changes its
energy. If the pomeron structure contains a significant quark
component, such events will be generated directly, by a quark emerging
from the pomeron, absorbing either a real or a  virtual photon, and in doing so
 being knocked
fast sideways.

The single power of $(1-x)$ in  the structure function (1) is characteristic
of a single spectator jet, which will be an antiquark if (1) represents
a quark structure function. In most of the high-$p_T$ events this
spectator will emerge along the beam direction, with another high-$p_T$
recoil jet generated by a hard QCD interaction. But there is also
the possibility that instead it is this spectator that is the recoil jet.
This is shown in figure 1. From this figure one can
see that such events would have a rather clean structure, with the
initial proton going down the beam pipe and otherwise rather like an
$e^+e^-$ event. All the energy of the pomeron goes into the pair of
transverse jets. (UA8 may already have some evidence that
this can happen\ref{5}.)
The calculation of figure 1 is the subject of this paper.

One could check that the proton has not broken up, and measure
how much energy it has lost to the pomeron, with an in-beam-pipe
detector. However, it is not essential to check this: one can allow
the proton to break up into a low-mass excited state, so
increasing the cross-section. All that is necessary\ref{8}
is that the fraction $\xi$ of the intial proton energy carried off by the
pomeron be at most a few percent, in order that one may be reasonably
sure that the pomeron has been radiated off from the proton, rather than,
for example, an $f$.  In principle, one can check that there is no
contamination from $f$ exchange, from the dependence of the cross-section
on $\xi$. But in practice values of $\xi$ rather less than 0.01 will
probably be studied, and then it should be safe to assume that there
is no contamination.

Since the pomeron takes only a small fraction of the initial proton
energy, one needs high energy such as HERA will provide in order
that this energy be enough to make a pair of high-$p_T$ jets. The
photon that is emitted by the electron and absorbed by the quark or antiquark
can be either real or virtual; we shall be content to calculate
the real-photon case. However, the $\gamma p$ cross-section for
the process we are considering should not vary rapidly with the $Q^2$
of the photon, so
that it will not be essential in the experiment to verify that the
photon is real, provided that it is reasonably sure that
$Q^2\ll P_T^2$.

We have previously adopted two different approaches to pomeron exchange.
One is very phenomenological and readily yields\ref{6} the form (1) for the
quark structure function. The other\ref{9} is based more on QCD: it
treats the pomeron exchange as approximately equivalent to the exchange
of two nonperturbative gluons. The success\ref{10} of this approach in
describing the exclusive process $\gamma ^* p \to \rho p$, particularly
in the light of
new data from the NMC collaboration\ref{2,11}, encourages us to use it here.

In the simplest approximation to pomeron exchange, we are thus
led to the four diagrams of figure 2, in which the two gluons couple
in different ways to the high-$p_T$ quarks. At their lower end, both
gluons couple to the same quark within the proton\ref{9}. We calculate
the imaginary part of the amplitude, which means that we do not include
diagrams in which the gluons cross each other. In this simple approximation
the amplitude is pure imaginary.

When we square the amplitude and calculate the cross-section, we need a
trace\footnote{$^*$}{We calculate the trace using the programme FORM
by J Vermaseren}. We retain only the leading terms in $1/\xi$ in this.
This means that when we impose the condition that the quark lines to which
the gluon with momentum $\ell$ is attached are on shell (because we
are calculating the imaginary part of the amplitude) we find that
in the case of the diagrams of figure 2a and 2b  we can
approximate  $\ell$ by
$$
\ell=-{{\ell _T^2}\over{\xi\; \hat t}}\; p + \ell _T
\eqno(2)
$$
where $\hat t=(q-P)^2$ and where the two-dimensional Euclidean
vector $\ell _T$ is transverse to $p$ and $P'$. For the other two diagrams
there is a similar expression, but with $\hat u=(q-P')^2$ replacing
$\hat t$ and with $\ell _T$ transverse to $P$ instead of $P'$.
In each term in the trace we need two momenta $\ell$ and $\ell '$, one for the
amplitude and one for its complex conjugate.
When we add together the contributions from all the diagrams the
terms in the trace that are independent of the $\ell _T$  and $\ell _T'$
cancel. Those
that are linear in either $\ell _T$ or $\ell _T'$ integrate to zero, and the
leading surviving terms are those proportional to
$\ell _T^2 \ell _T^{\prime 2}$.

At zero momentum transfer $t=(p-p')^2$, the diagrams of figure 2 together
give (for a real photon beam)
$$
{{d^3\sigma ^{\gamma p}}\over{dt d\xi dP_T^2}}=
     {{18\alpha\beta _0^4\mu _0^4\alpha _S(P_T^2)/\alpha _S^{(0)}}
\over{\pi ^2\xi ^2sP_T^4}}
{{1-2P_T^2/(\xi s)}\over{\sqrt{1-4P_T^2/(\xi s)}}}
\eqno(3)
$$
Here $\surd s$ is the $\gamma p$ centre-of-mass energy and we have used
our previous definition\ref{10} of the dimensionless
quantity $\beta _0 \mu _0$:
$$\int d\ell _T^2 \ell _T^2\; [\alpha _S^{(0)} D(-\ell _T^2)]^2=
  {{9\beta _0^2\mu _0^2}\over{8\pi}}
\eqno(4)
$$
where $-g^{\mu\nu}D$ is the Feynman-gauge propagator of the
nonperturbative gluon and $\alpha _S^{(0)}$ is its
coupling to an on-shell quark. We have found\ref{10} that experiment
requires $\alpha _S^{(0)} \approx 1$ and
$$
\beta _0^2\mu _0^2\approx 4
\eqno(5)
$$
The $\alpha _S(P_T^2)/\alpha _S^{(0)}$ appears in (3) because
one of the gluons is
coupled to a quark that goes far off shell at large $P_T^2$. There is the usual
ambiguity about what argument one should choose for $\alpha _S$ to account for
this, and we have not included any $K$-factor to account for higher-order
perturbative QCD corrections to the upper parts of the diagrams of figure 2.

In (3) we have summed the contributions from the three lightest
quarks and antiquarks. An interesting question is whether the pomeron
is flavour-blind
to the extent that, at large $P_T^2$, $c$-quark jets will be produced at
the same rate as $u$ quarks.

The form (3) is in the crudest model, where two gluons and only two
are supposed to model the pomeron exchange. In order to bring it into
contact with experiment, we have to refine the formula by replacing the
simple power $\xi ^{-2}$ with the  Regge power $\xi ^{-2\alpha (t)}$.
Here $\alpha (t)$ is the pomeron trajectory:
$$
\alpha (t) =1+\epsilon +\alpha 't
\eqno(6)
$$
where\ref{3} $\epsilon \approx 0.08$ and $\alpha '=0.25$ GeV$^{-2}$.
Also, when $t$ (the momentum transfer suffered by the initial proton)
is no longer zero, we need the proton's elastic form factor
$$
F_1(t)={{4m^2-2.8t}\over{4m^2-t}}{1\over{(1-t/0.7)^2}}
\eqno(7)
$$
Thus (3) becomes
$$
{{d^3\sigma ^{\gamma p}}\over{dt d\xi dP_T^2}}=\xi ^{-2\alpha (t)}
    [F_1(t)]^2 {{18\alpha\beta _0^4\mu _0^4\alpha _S(P_T^2)/\alpha _S^{(0)}}
\over{\pi ^2sP_T^4}}
{{1-2P_T^2/(\xi s)}\over{\sqrt{1-4P_T^2/(\xi s)}}}
\eqno(8)
$$

This formula should be valid so long as the momentum transfer $\hat t$
between the photon and the high-$p_T$ jet is large enough to get into
the perturbative region of the propagator of the exchanged quark. This
requires
$$
-\hat t \equiv \half\xi s\left (1-{\sqrt{1-4P_T^2/(\xi s)}}\right ) >- t_0
\eqno(9)
$$
where, for the light quarks, we guess that $|t_0|$ is about 1 GeV$^2$.
We have also assumed that $P_T$ is much larger than any masses,
and implicitly that it is large enough for there to be well-defined jets,
so we shall require $P_T$ to be at least 5 GeV. The fractional energy
$\xi$ of the pomeron must be greater than its lower kinematic limit
$4P_T^2/s$. For $P_T=5$ GeV jets produced at $\surd s=250$ GeV, this allows
$\xi$ to be extremely small: $\xi > .0016$. At higher energy than HERA,
where $\xi$ can be even smaller, ultimately it will be necessary to take
account of the possibility of replacing the pomeron in figure 1 with more than
one pomeron. This will reduce the effective negative power of $\xi$
in (8) and ultimately reduce it to a logarithm: compare
the total $pp$ or $\bar pp$ cross-section, which even at
Tevatron  energy behaves as a very-slowly-reducing power of $s$, but at
asymptotic energies we know that this power must give way to a behaviour
bounded by log$^2 s$ (the Foissart bound).

In order to estimate how large is the cross-section (9), approximate
the running coupling to be constant and equal to 0.2. Also, because
of the lower limit on $\xi$ we know that $1-2P_T^2/(\xi s)$ is
greater than $\half$, so we replace it with $\half$. We also approximate the
square root by unity, because if we integrate over $P_T^2$ the region where
it is not near unity gives only a small contribution.  Then
$$
{{d^3\sigma ^{\gamma p}}\over{dt d\xi dP_T^2}}\approx \xi ^{-2\alpha (t)}
    [F_1(t)]^2 {{0.02}
\over{sP_T^4}}
\eqno(10)
$$
and
$$
{{d^2\sigma ^{\gamma p}}\over{dt d\xi }}(P_T>P_{Tmin})\approx\xi ^{-2\alpha
(t)}
    [F_1(t)]^2 {{0.02}
\over{sP_{Tmin}^2}}
\eqno(11)
$$
It will be interesting to check this functional dependence on $\xi$.
However, if we integrate $\xi$ down to its kinematic limit then
$$
{{d\sigma ^{\gamma p}}\over{dt}}(P_T>P_{Tmin})\approx
{{(s/4P_{Tmin}^2)}^{2\alpha (t)-2}
    [F_1(t)]^2 {{0.005}
\over{(2\alpha (t)-1)P_{Tmin}^4}}}
\eqno(12)
$$
Finally, insert $[F_1(t)]^2 \approx e^{bt}$ with $b\approx 4$ GeV$^{-2}$
and use the fact that $\alpha (t)$ in (6) varies slowly with $t$;
then
$$
\sigma ^{\gamma p}(P_T>P_{Tmin})\approx
 {{0.005 (s/4P_{Tmin}^2)^{2\epsilon}}\over{P_{Tmin}^4(b+2\alpha
 '\log(s/4P_{Tmin}^2)}}
\eqno(13)
$$
For $\surd s=250$ GeV and $P_{Tmin}=5$ GeV this gives about
$10^{-33}$ cm$^2$. A more accurate numerical integration of (8), with the
same fixed value of $\alpha _S$, gives $1.4 \times 10^{-33}$ cm$^2$.

If we do not require the proton to remain intact, then to first approximation
we simply omit the elastic form factor\ref{12}, ie set $b=0$.
This doubles the cross section.
\vfill\eject
{\bf References}

\i{1} UA8 collaboration: P Schlein, LPHEP'91 Conference (Geneva, July 1991)

\i{2} P V Landshoff, LPHEP'91 Conference (Geneva, July 1991)

\i{3} A Donnachie and P V Landshoff, Nuclear Physics B267 (1986) 690

\i{4} G Ingelman and P Schlein, Physics Letters B152 (1985) 256

\i{5} UA8 collaboration: A Brandt, seminar at CERN, December 1991.

\i{6} A Donnachie and P V Landshoff, Nuclear Physics B303 (1988) 634

\i{7} H Fritzsch and K-H Streng, Physics Letters 169B (1985) 391; E~L~Berger,
J~C~Collins, D~E~Soper and G~Sterman, Nuclear Physics B286 (1987) 704;
N~Arteago-Romero, P~Kessler and J~Silva, Mod Phys Lett A1 (1986) 211

\i{8} A Donnachie and P V Landshoff, Nuclear Physics B244 (1984) 322

\i{9} P V Landshoff and O Nachtmann, Z Phys C35 (1987) 405

\i{10} A Donnachie and P V Landshoff, Nuclear Physics B311 (1988/9) 509

\i{11} NMC collaboration: P Amaudruz et al, preprint CERN-PPE/91-228

\i{12} J R Cudell, Nuclear Physics B336 (1990) 1

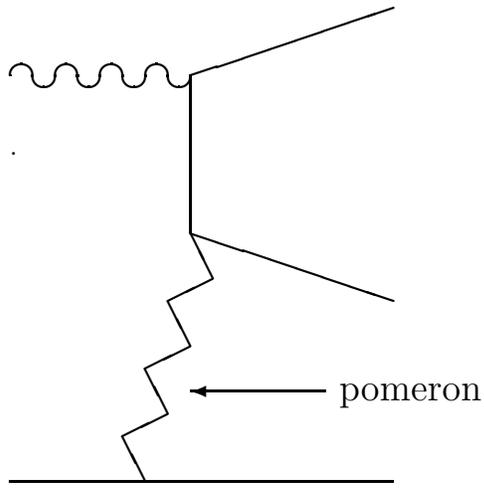
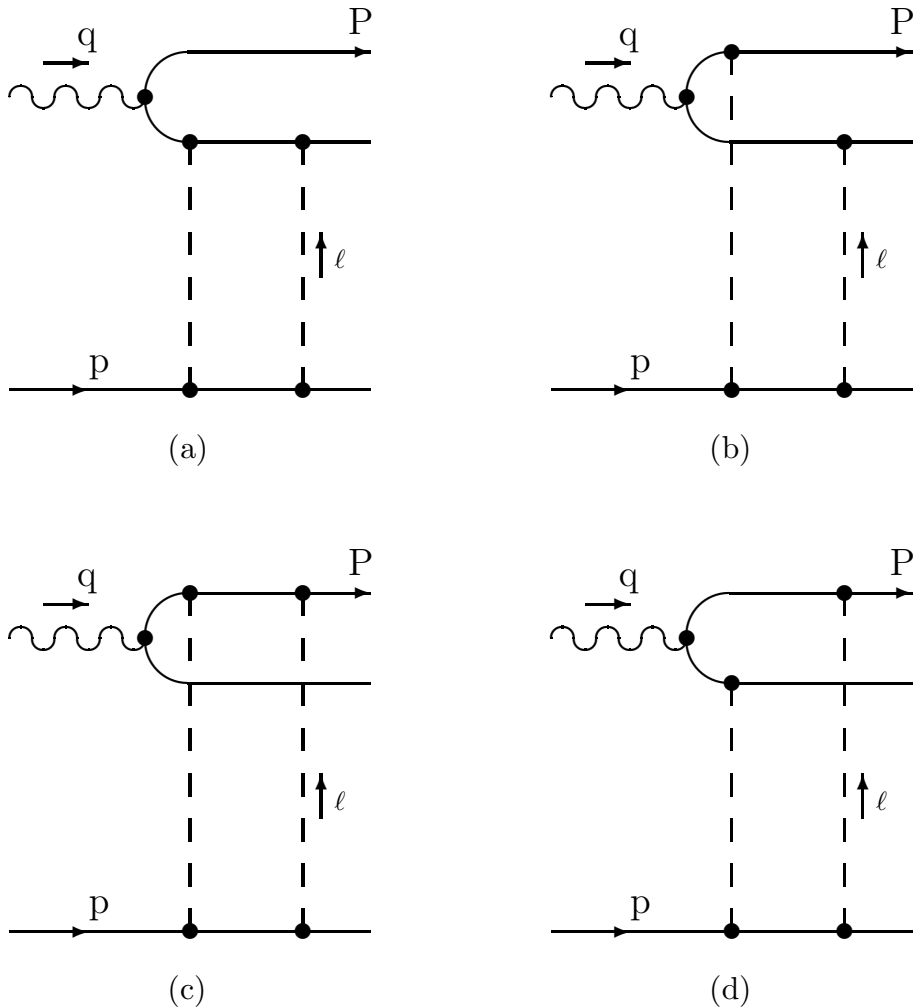
\bye